\documentclass[12pt]{iopart}
\usepackage{setstack}
\usepackage{graphicx}% Include figure files
\begin{document}
%--------------------------------------------------------------------
%\rightline{{\bf arXiv}: gr-qc/yymmnnn}
%-----------------------------------------------------------------
% Local definitions
\font\fiverm=cmr5\font\sevenrm=cmr7
\def\ns#1{_{\hbox{\sevenrm #1}}}
\def\dd{{\rm d}}\def\OM{\mean\Omega}\def\ab{\mean a}
\def\ds{\dd s} \def\e{{\rm e}} \def\etal{{\em et al}.}
\def\ga{\gamma}\def\de{\delta}
\def\th{\theta}\def\ph{\phi}\def\rh{\rho}\def\si{\sigma}

\def\mean#1{{\vphantom{\tilde#1}\bar#1}}
\def\bH{\mean H}\def\gb{\mean\ga}\def\gc{\gb\Z0}
\def\goesas{\mathop{\sim}\limits}

\def\w#1{\,\hbox{#1}} \def\Deriv#1#2#3{{#1#3\over#1#2}}
\def\Der#1#2{{#1\hphantom{#2}\over#1#2}} \def\br{\hfill\break}
\def\ts{t} \def\tc{\tau} \def\tv{\tc\ns{v}} \def\tw{\tc\ns{w}}
\def\Dtc{\mathop{\hbox{$\Der\dd\tc_{\!\lower1pt\hbox{\fiverm w}}$}}}

\def\Z#1{_{\lower2pt\hbox{$\scriptstyle#1$}}}
\def\X#1{_{\lower2pt\hbox{$\scriptscriptstyle#1$}}}
\def\av{{a\ns{v}\hskip-2pt}} \def\aw{{a\ns{w}\hskip-2.4pt}}
\def\etw{\eta\ns w} \def\etv{\eta\ns v}
\def\FF{{\cal F}} \def\Fi{\FF\X I}
\def\Hh{H} \def\Hm{H\Z0} \def\frn#1#2{{\textstyle{#1\over#2}}}
\def\half{\frn12} \def\DD{{\cal D}} \def\Vav{{\cal V}}
\def\lsim{\mathop{\hbox{${\lower3.8pt\hbox{$<$}}\atop{\raise0.2pt\hbox{$\sim$}}
$}}}
\def\kmsMpc{\w{km}\;\w{sec}^{-1}\w{Mpc}^{-1}}
\def\ab{{\bar a}} \def\LCDM{$\Lambda$CDM}
\def\OmMn{\Omega\Z{M0}}

\def\beq{\begin{equation}} \def\eeq{\end{equation}}
\def\bea{\begin{eqnarray}} \def\eea{\end{eqnarray}}
\def\ApJ#1{Astrophys.\ J.\ {\bf#1}} \def\PL#1{Phys.\ Lett.\ {\bf#1}}
\def\CQG#1{Class.\ Quantum Grav.\ {\bf#1}}
\def\GRG#1{Gen.\ Relativ.\ Grav.\ {\bf#1}}
%----------------------------------------------------------------
\title{Gravitational energy and cosmic acceleration}
%-----------------------------------------------------------------
\author{David L. Wiltshire}
%-----------------------------------------------------------------
\address{Department of Physics and Astronomy, University of Canterbury,
Private Bag 4800, Christchurch 8140, New Zealand}
%-----------------------------------------------------------------
\eads{\mailto{David.Wiltshire@canterbury.ac.nz}\\
http://www2.phys.canterbury.ac.nz/$\goesas$dlw24/}

\begin{abstract}
Cosmic acceleration is explained quantitatively, as an apparent effect due
to gravitational energy differences that arise in the decoupling of bound
systems from the global expansion of the universe. ``Dark energy'' is a
misidentification of those aspects of gravitational energy which by virtue
of the equivalence principle cannot be localised, namely gradients in the
energy due to the expansion of space and spatial curvature variations in
an inhomogeneous universe.
A new scheme for cosmological averaging is proposed which solves the
Sandage--de Vaucouleurs paradox. Concordance parameters fit supernovae
luminosity distances, the angular scale of the sound horizon in the CMB
anisotropies, and the effective comoving baryon acoustic oscillation scale
seen in galaxy clustering statistics.
Key observational anomalies are potentially resolved, and unique
predictions made, including a quantifiable variance in the Hubble flow
below the scale of apparent homogeneity.
\end{abstract}
%\pacs{98.80.-k, 98.80.Es, 98.80.Jk, 95.36.+x}
%Keywords: dark energy, theoretical cosmology, observational cosmology

\noindent{March 2007. An essay which received {\em Honorable Mention} in the
2007 Gravity Research Foundation Essay Competition.}
%-----------------------------------------------------------------
\maketitle
%-----------------------------------------------------------------
%\pageno=1
\section{Introduction}
%-----------------------------------------------------------------

Our most widely tested ``concordance model'' of the universe relies on the
assumption of an isotropic homogeneous geometry, in spite of the
fact that at the present epoch the observed universe is anything but smooth
on scales less than 150--300 Mpc. What we actually observe
is a foam--like structure, with clusters of galaxies strung in filaments
and bubbles surrounding huge voids. Recent surveys suggest that some 40--50\%
of the volume of the universe is in voids of a characteristic scale 30$h^{-1}$
Mpc, where $h$ is the dimensionless Hubble parameter, $\Hm=100h\kmsMpc$.
If larger supervoids and smaller minivoids are included, then it is fair to
say that our observed universe is presently void--dominated.

It is nonetheless true that a broadly isotropic Hubble flow is observed,
which means that a nearly smooth Friedmann--Lema\^{\i}tre--Robertson--Walker
(FLRW) geometry must be a good approximation at some level of averaging,
if our position is a typical one. In this essay, I will argue,
however, that in arriving at a model of the universe which is dominated
by a mysterious form of ``dark energy'' that violates the strong energy
condition, we have overlooked subtle physical properties
of general relativity in interpreting the relationship of our own
measurements to the average smooth geometry. In particular, ``dark
energy'' is a misidentification of those aspects of gravitational energy
which by virtue of the equivalence principle cannot be localized.

The proposed re--evaluation of cosmological measurements on the basis
of a universal {\em finite infinity} scale determined by primordial inflation,
leads to a new model for the universe. This model appears to pass key
observational tests, potentially resolves anomalies, and
makes new quantitative predictions.

\section{The fitting problem}

In an arbitrary inhomogeneous spacetime the rods and clocks of any set of
observers can only reliably measure local geometry. They give no indication
of measurements elsewhere or of the universe's global structure. By contrast,
in an isotropic homogeneous universe, where ideal observers are comoving
particles in a uniform fluid, measurements made locally are the same as those
made elsewhere on suitable time slices, on account of global symmetries.
Our own universe is somewhere between these two extremes.

By the evidence
of the cosmic microwave background (CMB) radiation,
the universe was very smooth at the time of last
scattering, and the assumption of isotropy and homogeneity was valid
then. At the present epoch we face a much more complicated fitting
problem \cite{fit1} in relating the geometry of the solar system, to that
of the galaxy, to that of the local group and the cluster it belongs to,
and so on up to the scale of the average observer in a cell which is
effectively homogeneous.

When we conventionally write down a FLRW metric
\beq\label{FLRW}
\dd\bar s^2 = - \dd\ts^2 +\ab^2(\ts)\dd\OM^2_{k}
\eeq
where $\dd\OM^2_{k}$ is the 3--metric of a space of constant curvature,
we ignore the fitting problem. In particular, even if the rods and clocks of
an ideal isotropic observer can be matched closely to the geometry
(\ref{FLRW}) at a volume--average position, there is no requirement of theory,
principle or observation that demands that such volume--average measurements
coincide with ours. The fact that we observe an almost isotropic CMB means
that other observers should also measure an almost isotropic CMB, if the
Copernican principle is assumed. However, it does not demand that other ideal
isotropic observers measure the same mean CMB temperature as us,
nor the same angular scale for the Doppler peaks in the anisotropy spectrum.
Significant differences can arise due to gradients in gravitational energy
and spatial curvature.

In general relativity space is dynamical and can carry energy and momentum.
By the strong equivalence principle, since the laws of physics must coincide
with those of special relativity at a point, it is only internal energy that
can be localized in an energy--momentum tensor on the r.h.s.\ of the
Einstein equations. Thus the uniquely
relativistic aspects of gravitational energy associated with spatial
curvature and geometrodynamics cannot be included in the energy momentum
tensor, but are at best described by a quasilocal formulation \cite{quasi}.

The l.h.s.\ of the Friedmann equation derived from (\ref{FLRW}) can be
regarded as the difference of a kinetic energy density per unit rest mass,
$E\ns{kin}=\half{\dot\ab^2\over\ab^2}$ and a total energy density per unit
rest mass $E\ns{tot}=-\half{k\over\ab^2}$ of the opposite sign to the
Gaussian curvature, $k$. Such terms represent forms of gravitational energy,
but since they are identical for all observers in an isotropic homogeneous
geometry, they are not often discussed in introductory cosmology texts. Such
discussions appear rarely in the cases of specific inhomogeneous models,
such as the Lema\^{\i}tre--Tolman--Bondi (LTB) solutions.

In an inhomogeneous cosmology, gradients in the kinetic energy of expansion
and in spatial curvature, will be manifest in the Einstein tensor, leading
to variations in gravitational energy that cannot be localized. The
observation that space is not expanding within bound systems implies that a
kinetic energy gradient must exist between bound systems and the volume
average in expanding space.
Furthermore, the fact that space within galaxies is well approximated by
asymptotically flat geometries implies that if there is significant
spatial curvature within our present horizon volume, then a spatial
curvature gradient should also contribute to the gravitational energy
difference between bound systems and the volume average.

\section{Finite infinity and boundary conditions from primordial inflation}

In his pioneering work on the fitting problem, Ellis \cite{fit1}
suggested the notion of {\em finite infinity}, ``{\em fi}$\,$'',
as being a timelike surface within which the dynamics of an isolated
system such as the solar system can be treated without reference to the
rest of the universe. Within finite infinity spatial geometry might be
considered to be effectively asymptotically flat, and governed by ``almost''
Killing vectors.

Quasilocal gravitational energy is generally defined in terms of surface
integrals with respect to surfaces of a fiducial spacetime, and for the
discussions of binding energy and rotational energy to which the quasilocal
approach is commonly applied, asymptotic flatness
is usually assumed. I propose that to quantify cosmological gravitational
energy with respect to observers in bound systems an appropriate notion
of finite infinity must be used as the fiducial reference point, since
bound systems can be considered to be almost asymptotically flat.

To date Ellis' 1984 suggestion \cite{fit1} has not been further developed,
perhaps because there is no obvious way to define finite infinity
in an arbitrary inhomogeneous background. To proceed I will make the crucial
observation that since our universe was effectively homogeneous and
isotropic at last scattering, a notion of a universal critical density
scale did exist then. It was the density required for gravity
to overcome the initial uniform expansion velocity of the dust fluid.
I will assume, as consistent with primordial inflation, that the present
horizon volume of the universe was very close to the critical density
at last scattering, with scale--invariant perturbations.

Since the evolution of inhomogeneities involves back--reaction we must use
an averaging scheme such as that developed by Buchert \cite{buch1}. An
important lesson of such schemes is that averaging a
quantity such as the density, and then evolving the average by
the Friedmann equation, is not the same as evolving the inhomogeneous Einstein
equations and then taking the average. Thus even if our present horizon
volume, $\cal H$, was close to critical density at last scattering,
differing perhaps by a factor of $\left.\de\rh/\rh\right|\Z{{\cal H}i}\goesas
-10^{-5}$, the present horizon volume can nonetheless have a density
well below critical. Furthermore,
the present {\em true critical density} or {\em closure density}
which demarcates a bound system from an unbound
region, can be very different from the notional critical density
inferred from a FLRW model using the presently measured global Hubble
constant, $\Hm$. This circumstance can in fact be understood as outcome of
cosmic variance combined with the scale--invariance of the primordial
perturbation spectrum resulting from inflation, and the subsequent causal
evolution of such inhomogeneities \cite{opus}.

In ref.\ \cite{opus} I provide a technical definition of finite infinity
in terms of the evolution of the true critical density. Finite infinity
represents an averaging scale with a non--static boundary analogous to
the spheres cut out in the Einstein--Straus Swiss cheese model \cite{gruyere},
but it involves average geometry rather
than matching exact solutions, and no assumptions about homogeneity
are made outside finite infinity. Finite infinity represents a physical
scale expected to lie outside virialized galaxy clusters, but within the
filamentary walls surrounding voids. Since it is a scale related to
the true critical density, space at finite infinity boundaries can be
described by the spatially flat metric
\beq\ds^2\Z{\Fi}=-\dd\tw^2+\aw^2(\tw)\left[\dd\etw^2+
\etw^2\left(\dd\th^2+\sin^2\th\,\dd\phi^2\right)\right]\,.
\label{figeom}\eeq

Beyond finite infinity, the spatial geometry is not given by (\ref{figeom}).
Since we live in a universe dominated by voids at the present epoch, a
two--scale approximation can be developed by assuming that the geometry
near the centres of voids is given by
\beq\ds^2\Z{\DD\X C}=-\dd\tv^2+\av^2(\tv)\left[\dd\etv^2+\sinh^2(\etv)
\left(\dd\th^2+\sin^2\th\,\dd\phi^2\right)\right]\,,
\label{vogeom}\eeq
where the local spatial curvature is negative, differing from that of
(\ref{figeom}) determined by observers in galaxies within finite infinity.
Furthermore the local void time parameter, $\tv$, differs from that
within finite infinity regions, on account of gravitational energy
differences. Clocks run slower where mass is concentrated, but because
this time dilation relates mainly to energy associated with spatial curvature
gradients, the differences can be significantly
larger than those we would arrive at in considering only binding energy
below the finite infinity scale, which is very small.

In ref.\ \cite{opus} Buchert's scheme is applied to the evolution of a
volume average of the two geometries (\ref{figeom}) and (\ref{vogeom}).
The average spatial geometry does not have a simple uniform
Gaussian curvature, but can be described in terms of an
effective scale factor $\ab(t)\equiv\left[\Vav(t)/\Vav_i\right]^{1/3}$
related to the evolution of the spatial volume, $\Vav$, over a suitable
averaging scale, which I identify as the time evolution of present
particle horizon volume. The volume--average geometry which
replaces the metric (\ref{FLRW}) of a FLRW universe will also have a time
parameter, $t$, which differs from that measured by observers within finite
infinity regions, via a mean lapse function
\beq \dd t=\gb(\tw)\,\dd\tw\,. \eeq
Effectively, $\tw$, is a universal cosmic time set by the almost stationary
Killing vector of the finite infinity scale, which differs from the local
time of a volume--average observer. Such a volume average observer
would measure an older age of the universe, and a lower mean CMB
temperature. Although this may seem surprising it is entirely consistent
with observation, since we exchange photons with other bound systems which
keep a time close to the universal finite infinity time scale, if
binding energy is neglected. At early times, $\gb\simeq1$.
It grows monotonically, reaching values of order $\gc\goesas1.38$ today.

The observation of an isotropic Hubble flow is satisfied by adopting
a uniform expansion gauge, {\em when expansion is referred to local
measurements}, as a change of proper length with respect to
proper time. Referred to any one set of clocks, it appears that voids
expand faster than the filamentary bubble walls where galaxy clusters
are located. Nonetheless if we take account of the fact that clocks
tick faster in voids, the locally measured expansion can still be
uniform. This provides an implicit resolution of the Sandage--de Vaucouleurs
paradox: in the standard FLRW paradigm, the statistical scatter in
the Hubble flow should be so large that no Hubble constant can be
extracted below the scale of homogeneity. Yet Hubble originally derived
a linear law on scales of 20 Mpc, of order 10\% of the scale of apparent
homogeneity.

While dark energy has been invoked in a qualitative way to explain the
Sandage--de Vaucouleurs paradox, quantitative attempts to resolve the paradox
have not been fully successful in the \LCDM\ paradigm.
The new model universe \cite{opus} makes a quantitative prediction as
to the variance of the Hubble flow below the scale of apparent homogeneity.
The Hubble parameter observed within filamentary walls,
$\bH=\gb^{-1}\Hh+\gb^{-2}\Dtc\gb$, is lower than the global average Hubble
parameter, $\Hh$. Similarly, a Hubble parameter larger than the global
average will be observed across the nearest large voids of diameter
30$h^{-1}$ Mpc. Since voids occupy a greater volume of space than bubble
walls, an isotropic average over small redshifts will give an overall
higher Hubble constant locally until the scale of apparent homogeneity is
reached, when we sample the global average fractions of walls and
voids. This is consistent with the observed ``Hubble bubble''
feature \cite{JRK,essence}.

\section{Apparent cosmic acceleration without ``dark energy''}

The volume--average Buchert equations for the two scale model have
been integrated\footnote{{\em Note added}: Numerical integrations were
initially undertaken in ref.\ \cite{opus}. However, after this essay was
submitted an exact solution of the two--scale Buchert equations was
subsequently obtained \cite{sol}.} in ref.\ \cite{opus}, and best--fit
parameters for initial conditions consistent with the evidence of the CMB
radiation, and the expectations of primordial inflation, are provided in
ref.\ \cite{LNW}.

The present model provides a definitive quantitative answer to the
debate about whether back--reaction can mimic cosmic
acceleration \cite{camp1,camp2}. It is found that, as measured by a
volume--average observer the expansion appears to decelerate, albeit with a
deceleration parameter close to zero. This would vindicate the claims of
those who have argued that back--reaction is too small to be a source of
cosmic acceleration \cite{camp2}, if the position of the observer
could be neglected. However, we are in a bound system, not expanding
space, and observers in galaxies whose clocks tick slower than the volume
average can nonetheless still register apparent acceleration.

Gravitational energy and spatial curvature gradients between bound systems
and the volume--average position in a void--dominated universe --
the intrinsic physics of a curved expanding space and its
affect on measurements -- are therefore the essential ingredients to
understanding apparent cosmic acceleration. By overlooking the
operational basis of measurements in general relativity, we have
come to misidentify gravitational energy gradients as ``dark energy''.

The coincidence as to why cosmic
``acceleration'' should occur at the same epoch when the largest structures
form is naturally solved. Voids are associated with negative spatial
curvature, and negative spatial curvature is associated with the positive
gravitational energy which is largely responsible for the
gradient between bound systems and the volume average. Since
gravitational energy directly affects relative clock rates, it is at the
epoch when the gravitational energy gradient changes significantly that
apparent cosmic acceleration is seen.

In the new paradigm, apparent cosmic acceleration, which occurs for
redshifts of order $z\lsim0.9$, is less extreme than in the \LCDM\ paradigm,
and the universe is very close to a coasting Milne universe at late
times, in accord with observation. Fits to the Riess06 gold data
set \cite{Riess06} yield values of $\chi^2\simeq0.9$ per degree of
freedom \cite{opus,LNW}, while a Bayesian model comparison indicates
that the results statistically indistinguishable from the \LCDM\
model \cite{LNW}. As shown in Fig.\ 1, parameters can be
found which simultaneously fit supernovae, the angular scale of the sound
horizon which sets the angular scale of the CMB Doppler peaks, and
the effective ``comoving'' scale of the baryon acoustic oscillation as
measured in galaxy clustering statistics \cite{bao}. It is remarkable
that these values agree precisely with the value of the Hubble
constant recently determined by the HST Key Team of Sandage
\etal \cite{Sandage}, as this value differs by 14\% from that claimed as
a best--fit to the \LCDM\ paradigm with WMAP \cite{wmap}.

The new paradigm may also resolve observational anomalies.
The expansion age is larger, allowing more time for structure formation.
The universe is typically about 14.7 Gyr old as viewed from a
galaxy, or 18.6 Gyr at the volume average.

Since the baryon--to--photon ratio is conventionally defined at the
volume average, a systematic recalibration of
cosmological parameters is required. Using standard big bang nucleosynthesis
bounds, a best--fit ratio of non--baryonic matter to baryonic matter of 3:1
is found. Non--baryonic dark matter is still very significant, but reduced
relative to the \LCDM\ paradigm, making it possible to have enough baryon drag
to fit the ratio of heights of the first two Doppler peaks, while
simultaneously better fitting helium abundances and potentially resolving
the lithium abundance anomaly \cite{lithium}.

\begin{figure}[htb]
\vbox{
\centerline{{\bf(a)}\scalebox{0.75}{\includegraphics{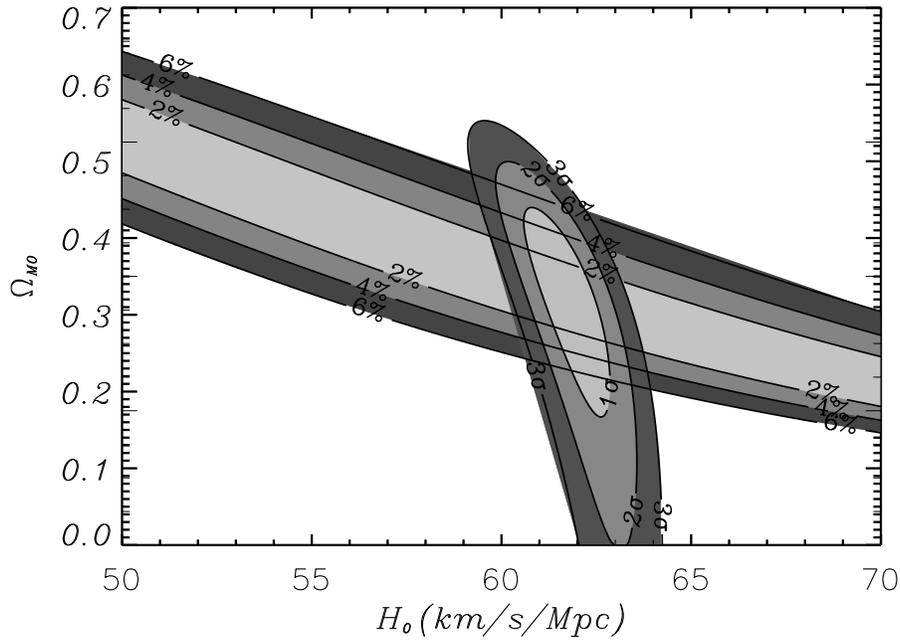}}}
\centerline{{\bf(b)}\scalebox{0.75}{\includegraphics{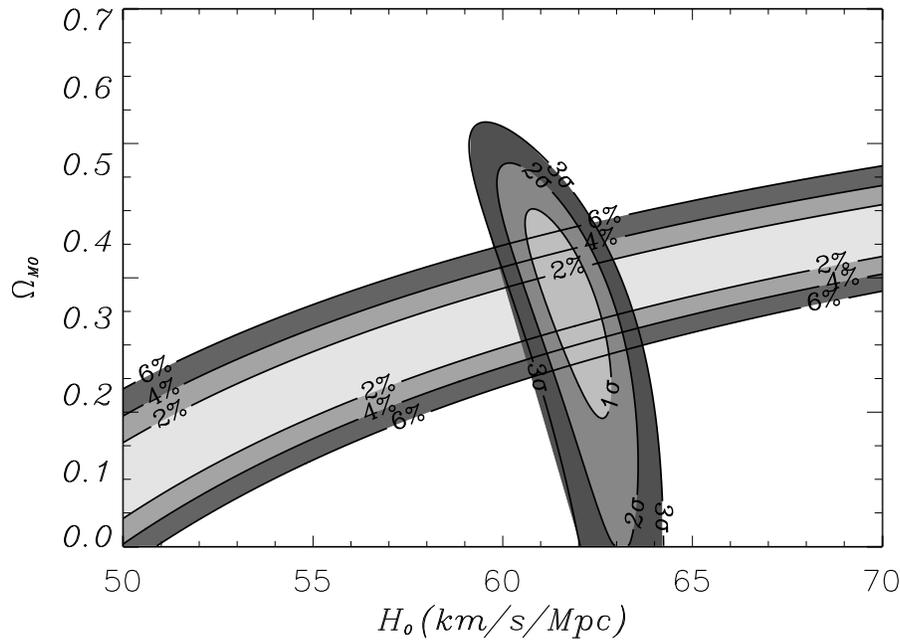}}}
%\centerline{{\bf(a)}{\includegraphics[width=125truemm]{dL_ang.png}}}
%\centerline{{\bf(b)}{\includegraphics[width=125truemm]{dL_BAO.png}}}

\caption{\label{fig}%
{\sl In panel {\bf(a)} 1$\si$, 2$\si$ and 3$\si$ confidence limits (oval)
for fits of luminosity distances of type Ia supernovae (SneIa) in the Riess06
gold dataset \cite{Riess06} are compared to parameters within the
($\OmMn$,$\Hm$) plane which fit the angular scale of the sound horizon
$\de=0.01$ rad deduced for WMAP \cite{wmap}, to within 2\%, 4\% and 6\%.
In panel {\bf(b)} the SneIa are similarly compared to an effective comoving
baryon acoustic oscillation (BAO) scale of $104h^{-1}$Mpc, as verified in
galaxy clustering statistics \cite{bao}. The sound horizon angular scale
and BAO comoving scale generally fit for different parameters, yet agree with
each other for the same parameters that best fit the SneIa
data, in a range that also agrees with the Hubble constant measurement of
Sandage \etal\ \cite{Sandage}: $\Hm=62.3\pm1.3\w{(stat)}\pm5.0\w{(syst)}
\kmsMpc$. We find $\Hm=61.7^{+1.2}_{-1.1}\kmsMpc$ and
$\OmMn=0.33^{+0.11}_{-0.16}$ using $1\si$ statistical uncertainties
from SneIa only. A 4\% match to both sound horizon and BAO scales
would reduce the bounds on $\OmMn$ by a factor of two. Note: $\OmMn$ is
the ``conventional dressed density parameter''\cite{opus}.
Further details are provided in ref.\ \cite{LNW}.}}}
\end{figure}

The angular scale of the Doppler peaks is often claimed to be a ``measure
of spatial curvature'', but that is only true in the FLRW paradigm, when
spatial curvature is assumed to be the same everywhere. In the new
paradigm, the angular scale might be claimed as a measure of {\em local}
spatial curvature in the 24\% of our present epoch horizon volume occupied
by the filamentary bubble walls, where galaxies are located. As Fig.\
\ref{fig} demonstrates, the angular scale can still be fit despite average
negative spatial curvature at the present epoch. This in fact may also
resolve the anomaly associated with ellipticity in the CMB anisotropy
spectrum \cite{ellipticity}, the observation of which implies the greater
geodesic mixing associated with average negative spatial curvature.
The question of other anomalies, and several possible directions of
research are discussed at length in ref.\ \cite{opus}.

\section{Conclusion}
%-----------------------------------------------------------------

Cosmic acceleration can be quantitatively explained within general
relativity as an apparent effect due to gravitational energy differences that
arise in the decoupling of bound systems from the global expansion of the
universe, without any exotic ``dark energy''. We must account not only for
the back--reaction of inhomogeneities in the Einstein equations, but also
for the fact that as observers in bound systems our measurements can differ
systematically from those at the volume average. This entails understanding
the subtle aspects of gravitational energy that exist by virtue of the
equivalence principle, the dynamical nature of general relativity and
boundary conditions from primordial inflation. The
results of refs.\ \cite{opus,LNW} show that it is likely that a
viable concordance model of the universe will be found. Detailed modelling
will not only rely heavily on new observational data, but will also be an
adventure into as yet still largely unexplored theoretical aspects of
general relativity.

The revolution that Einstein began exactly 100 years ago, when
he first thought about the equivalence principle in 1907, is not yet over. It
remains to us, the generation that has had the first real glimpse of what the
universe actually looks like, to think equally deeply about the operational
issues surrounding measurements, and the conceptual basis of general
relativity, in its application to the universe as a whole.

%------------------------------------------------------------
\ack
%------------------------------------------------
This work was supported by the Mardsen Fund of the Royal Society of
New Zealand.

\section*{References}
%-----------------------------------------------------------------

%-----------------------------------------------------------------
\end{document}